\documentclass{mem}
\usepackage{natbib}
\usepackage{txfonts}
\usepackage{balance}
\usepackage{graphicx}
\idline{75}{282}
\begin{document}

\title{Lithium destruction and production observed in red giant
  stars\thanks{Based on observations at the Very Large Telescope of the
    European Southern Observatory, Cerro Paranal/Chile under Programme
    083.D-0046(A).}}
%   \subtitle{trialtitle}

\author{
S.\,Uttenthaler\inst{1}
\and
T.\,Lebzelter\inst{1}
\and
M.\,Busso\inst{2}
\and
S.\,Palmerini\inst{3,4}
\and
B.\,Aringer\inst{1}
\and
M.\,Schultheis\inst{5}
}
%\offprints{Uttenthaler}

\institute{
University of Vienna, Department of Astrophysics, T\"urkenschanzstra\ss e 17,
1180 Vienna, Austria. \email{stefan.uttenthaler@univie.ac.at}
\and
Dipartimento di Fisica, Universit\`{a} di Perugia, and INFN, Sezione di
Perugia. Via Pascoli, 06123 Perugia, Italy
\and
Centro Siciliano di Fisica Nucleare e Struttura dalla Materia, Viale A.\ Doria,
6 - 95125 Catania, Italy
\and
Laboratori Nazionali del Sud I.N.F.N., via S.\ Sofia 62, Catania, Italy
\and
Institut Utinam, CNRS UMR6213, OSU THETA, Universit\'{e} de Franche-Comt\'{e},
41bis avenue de l'Observatoire, 25000 Besan\c{c}on, France
}

\authorrunning{Uttenthaler et al.}

\titlerunning{Li in red giants}

\abstract{
According to standard stellar evolution, lithium is destroyed throughout most of
the evolution of low- to intermediate-mass stars. However, a number of evolved
stars on the red giant branch (RGB) and the asymptotic giant branch (AGB) are
known to contain a considerable amount of Li, whose origin is not always
understood well. Here we present the latest development on the observational
side to obtain a better understanding of Li-rich K giants (RGB), moderately
Li-rich low-mass stars on the AGB, as well as very Li-rich intermediate-mass AGB
stars possibly undergoing the standard hot bottom burning phase. These last ones
probably also enrich the interstellar medium with freshly produced Li.
}
\maketitle{}

\section{Introduction}

Lithium (Li) is not only important for testing big bang nucleosynthesis
predictions and for studying diffusion processes in atmospheres of dwarf stars
\citep[see e.g.][and contributions in this issue]{Korn06}, it is also an
important diagnostic tool for stellar evolution. Its abundance strongly depends
on the ambient conditions because it is quickly destroyed at $T>3\times10^6$\,K
so that it diminishes if the stellar surface is brought into contact with hot
layers by mixing processes.

%In low-mass stars during the main-sequence phase, Li strongly decreases from
%its initial abundance, hence slow mixing should prevail \citep{Mic86,Pal03}.
%During the ascent on the red giant branch (RGB), any Li remaining in the
%envelope is further diluted by the first dredge-up (FDU); stellar models
%excluding atomic diffusion and rotation predict a surface Li abundance at FDU
%of $\log\epsilon({\rm Li})\le1.5$ at this stage. This is indeed what is
%observed in most G-K giants \citep{Lam80,Bro89,Mis06}, where some of these
%stars show Li abundances even far below the expectations \citep{Mal99}.

A nice illustration of the evolution of the Li abundance in low-mass,
low-metallicity stars during their ascent on the RGB can be found in Fig.~5 of
\citet{Lind09}, who investigated the globular cluster \object{NGC 6397}. The
least evolved main sequence and turn-off stars define a plateau of Li abundances
that is in agreement with the well-known Spite plateau \citep{Spite82}. A sharp
drop in Li abundance occurs once the stars undergo the first dredge-up (FDU) on
the sub-giant branch: the abundance drops asymptotically by more than an order
of magnitude towards a final value of $\log\epsilon({\rm Li})\approx1.1$, in
agreement with the maximum value of 1.5 predicted by stellar models excluding
atomic diffusion and rotation. This is also observed in most G-K giants in the
field \citep{Lam80,Bro89,Mis06}, where some of these stars show Li abundances
even far below the expectations \citep{Mal99}. During FDU, the convective
envelope is diluted by H-processed material that is devoid of Li. When the
outward advancing H-burning shell reaches the discontinuity in mean
mo\-le\-cu\-lar weight ($\mu$ barrier) left behind by FDU, the star is at the
bump in the luminosity function (RGB bump), and the Li abundance is further
diminished on the star's surface. This happens because the radiative layer
between the H-burning shell and the convective envelope is now chemically
homogeneous, so that any instability in the layer can lead to slow mixing
processes \citep[e.g.\ thermohaline mixing;][]{CZ07} that brings the remaining
Li to hot layers where it is burned. In the cluster investigated by
\citet{Lind09}, the Li abundance in some stars evolved beyond the RGB bump drops
to below the detection threshold at $\log\epsilon({\rm Li})\sim0.2$. For more
massive stars, the H-burning shell does not cross the $\mu$ barrier before it
reaches the early AGB, when again slow mixing processes can further diminish the
surface Li abundance.

However, about 1 -- 2\,\% of the K giants have a Li abundance higher than this
classical limit of $\log\epsilon({\rm Li})\approx1.5$ and are therefore called
{\em Li-rich} \citep{WS82,Bro89,dlR97}. Because it is known that these stars
underwent normal FDU, it is commonly believed that these stars cannot have
retained their original Li abundance, but rather must have replenished it
somehow in situ. Li destruction can turn into production if the overturn time
scale for mixing between the H-burning shell and the convective envelope becomes
faster than the decay of the parent nucleus $^7$Be. This is known as the
Cameron-Fowler mechanism
\citep[$^3$He($\alpha$,$\gamma$)$^7$Be(e$^-$,$\nu$)$^7$Li;][]{CF71}.

In this contribution we review our observational results of the Li destruction
and production in red giant stars with various masses and in various
evolutionary stages. We focus on three different topics here: first we present
our results from a recent spectroscopic survey of Li in RGB stars in the
Galactic bulge \citep{Leb12}, second we review the correlation between Li and
the third dredge-up indicator technetium (Tc) in low-mass, oxygen-rich AGB stars
\citep{Utt07,UL10,Utt11}, and finally we report on the large abundance of Li
detected in long-period Miras that probably undergo a phase of hot bottom
burning.

\section{A lithium survey in Galactic bulge RGB stars}

\subsection{Motivation}

There were two motivations for us to conduct a survey of the Li abundance along
the RGB in the Galactic bulge. The first one was the discovery of Li-rich AGB
stars in the outer Galactic bulge by \citet{Utt07}. These stars were interpreted
to be a result of fast mixing below the convective envelope called
``cool bottom processing'' \citep{SB99} so that the Cameron-Fowler mechanism is
activated. With a survey of the Li abundances in RGB stars in the bulge we aimed
at checking whether a surplus of Li is already present in this earlier
evolutionary stage, or if it must have been produced on the upper AGB.
%A lack of Li-rich
%stars on the RGB would be a clear sign that the Li found in the AGB stars must
%be produced in this late stage of stellar evolution. However, the presence of
%Li-rich stars on the RGB at a percentage comparable to that on the AGB would
%not allow for a clear interpretation because the giant branch below the RGB tip
%is a mixture of H-shell burning (RGB) stars and He-shell burning (early AGB)
%stars.

The second motivation was the claim by \citet{CB00} that the mentioned Li-rich
K giants are concentrated in two regions in the HR diagram: around the RGB bump
region of low-mass stars (those which undergo a He-core flash at the RGB tip),
and around the red clump/early AGB of intermediate-mass stars. In both these
phases, in the absence of a $\mu$ barrier fast extra-mixing could connect the
$^{3}$He-rich convective envelope with the H-burning shell, enabling fresh Li to
be produced by the Cameron-Fowler mechanism. This Li-rich phase would be
extremely short-lived, explaining the low number of Li-rich giants observed.
However, most of the hitherto investigated samples were either inhomogeneous in
mass or they were too small to verify these claimed distinct Li-rich episodes.
Our aim was to provide a check of these claims with a large, homogeneous sample
of RGB stars. The bulge contains a huge number of low-mass ($\sim1.1M_{\sun}$)
RGB stars at roughly equal distance, hence well-constrained luminosity.

\subsection{Target selection, observations, and data analysis}

\begin{figure}[t!]
\resizebox{\hsize}{!}{\includegraphics[clip=true]{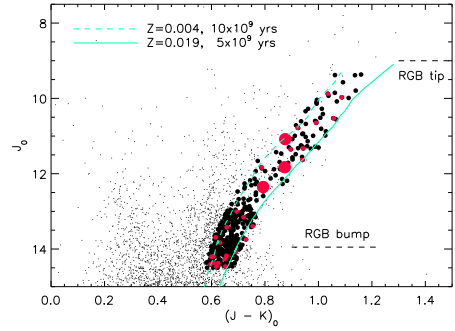}}
\caption{\footnotesize
CMD of the centre of the PG3 field, a 25\arcmin\ diameter field towards
$(l,b)=(0\degr,-10\degr)$. Big black dots represent stars that were observed in
this survey, small red circles represent Li-{\em detected} stars, whereas large
red circles represent Li-{\em rich} stars.}
\label{CMD}
\end{figure}

We selected the spectroscopic targets from a 2MASS colour-magnitude diagram
(CMD) in a 25\arcmin\ diameter circle towards the direction
$(l,b)=(0\degr,-10\degr)$, which is the centre of the Palomar-Groningen field
no.~3 (PG3). This field was chosen because also the sample of AGB stars studied
by \citet{Utt07} is located in the PG3. The CMD of this field with an
illustration of the target selection is displayed in Fig.~\ref{CMD}. Targets
were chosen along the RGB of two isochrones from \citet{Gir00} with ages and
metallicities as indicated in the legend. Furthermore, only stars fainter than
the RGB tip brightness (about 9\fm0 in $J_0$) were selected to avoid AGB stars,
and a minimum brightness of $J_0\ge14\fm5$ was demanded as to include also
stars at or slightly below the expected RGB bump, but to keep the exposure times
within reasonable limits.

%+++++++++++++++++++++++++++++++++++++++++++++++++++++++++++++++++++++++++++++++
%+++++++++++++++++++++++++++++++++++++++++++++++++++++++++++++++++++++++++++++++

The targets were observed with the FLAMES spectrograph at the VLT in the HR15
setting that covers also the Li\,I 671\,nm resonance line. The resolving power
of the spectra is 17\,000. Of the 514 targets that were initially proposed for
observations, 401 had spectra of sufficient quality to investigate the Li line.

The spectra were analysed with the help of COMARCS model atmospheres and
spectral synthesis \citep{Ari09}. The stellar effective temperature was
derived from the $(J-K)_0$ colour using the calibration relation established
from a series of COMARCS model atmospheres \citep{Leb12}. The logarithmic
surface gravity $\log g$ was derived from the isochrones in Fig.~\ref{CMD},
assuming that the stars belong to the bulge RGB. With the main stellar
parameters fixed in this way, we calculated a series of model atmospheres
with a range of metallicities and synthetic spectra based on these models. The
metallicity of each star was estimated by interpolating between the synthetic
spectra, using a $\chi^2$ minimisation scheme.

With the stellar parameters fixed in this way, we synthesised spectra, assuming
varying Li abundances to fit the observed spectra. Lithium was detected in 30
stars, three of which are Li-{\em rich} according to the criterion established
in the Introduction ($\log \epsilon({\rm Li})>1.5$), and 27 are
Li-{\em detected} ($\log \epsilon({\rm Li})\le1.5$). The most Li-rich sample
star has a Li abundance at the level of $\log \epsilon({\rm Li})=3.2$, in
agreement with the cosmic value, and should therefore be regarded as {\em super}
Li-rich. A comparison of spectra of a Li-rich star, a Li-detected star, as well
as a Li-poor star is presented if Fig.~2 of \citet{Leb12}.
%Figure~\ref{is42} shows the observed spectrum of that star with a
%synthetic spectrum fitted to it, along with spectra of a Li-detected star and a
%Li-poor star, all of them with similar stellar parameters. 
%
%%++++++++++++++++++++++++++++++++++++++++++++++++++++++++++++++++++++++++++++++
%\begin{figure}[t!]
%\resizebox{\hsize}{!}{\includegraphics[clip=true]{../LiRGB/tatltuae.ps}}
%\caption{\footnotesize
%Observed spectra of the the programme stars \#042, the most Li-rich star in our
%sample; \#030, a Li-detected star; and \#051, a Li-poor star (black dots, from
%bottom to top). For clarity, the spectra of stars \#030 and \#051 have been
%shifted by +0.2 and +0.4 in flux. The continuous line is the best-fit synthetic
%spectrum to star \#042 with a Li abundance of $\log\epsilon({\rm Li})=+3.2$.
%The two dotted vertical lines indicate the laboratory wavelengths of the
%hyperfine transitions in $^7$Li. Note also the good fit to two Fe lines at
%\mbox{$\sim670.55$} and $\sim671.21$\,nm, which indicates a well-determined
%me\-tal\-li\-ci\-ty of $[{\rm M}/{\rm H}]=-0.85$ for star \#042. The
%metallicities for the two other stars were determined to be $-0.94$ (\#030) and
%$-1.03$ (\#051), respectively.}
%\label{is42}
%\end{figure}
%%++++++++++++++++++++++++++++++++++++++++++++++++++++++++++++++++++++++++++++++

\subsection{Results}

Figure~\ref{CMD} also summarises the main results of this survey. The
Li-detected stars distribute all along the RGB, from below the RGB bump up to
the tip. We cannot discern a distinct episode of Li enrichment, neither at the
RGB bump luminosity nor anywhere else. The three Li-rich stars are clearly
brighter than the RGB bump and the red clump: the faintest of them is
$\sim1\fm4$ brighter than the bump, and $\sim0\fm7$ brighter than the bright
red clump \citep[note that there are two red clumps present in the outer
Galactic bulge, see][]{Nat10,MZ10}. If the Li-rich stars are genuine Galactic
bulge stars, and there is no reason to doubt this, then they are connected to
neither the RGB bump nor the red clump phase of evolution.

This casts some doubt on the claim by \citet{CB00} that a distinct phase of Li
enrichment occurs in the evolution of low-mass stars at the RGB bump. There are
a number of studies that draw similar conclusions. \citet{Gon09} identify
13 bulge stars with a detectable Li line, two of which are Li-rich, among a
sample of $\sim400$ stars. Their sample stars are $\sim0\fm7$ brighter than the
horizontal branch red clump. If these stars are indeed connected to the RGB bump
phase, they would more likely belong to the foreground disc, at a distance of
$\sim4.6$\,kpc. \citet{Mon11} searched their sample of 824 thick disc candidates
for the presence of Li, and found five Li-rich stars. Although only
spectroscopic distance estimates are available for these stars, they clearly
distribute all along the RGB, not connected to any distinct evolutionary phase.
\citet{Alc11} report on the detection of a Li-rich, low-mass M giant close to
the RGB tip. Finally, \citet{Ruch11} report the discovery of a number of
metal-poor, Li-rich giants distributed in a large range of luminosities, only
some of which might be connected to the RGB bump. There is growing evidence
that argues against a distinct Li-rich episode connected to the RGB bump of
low-mass red giants. Other parameters such as magnetic fields and the presence
of planets might also play an important role \citep{Gon10}.

On the other hand, \citet{Kumar11} searched 2000 giants in the Galactic disc,
and find that the Li-rich objects among them are connected to the red clump.
This result can be reconciled with the evidence discussed in the previous
paragraph if the stars in the sample of \citet{Kumar11} are on average more
massive than the stars in the other samples. For instance, bulge giants have a
typical mass of $1.1M_{\sun}$, whereas most Li-rich stars in the sample of
\citet{Kumar11} have masses $>1.6M_{\sun}$. In fact, also \citet{Kumar11} do not
find stars of $\sim1.0M_{\sun}$ that are clearly connected to either the RGB bump
or red clump phase (see their Fig.~2).

\citet{Utt07} identified four AGB stars in the PG3 field with Li abundance
of $\log\epsilon({\rm Li})=0.8$, 0.8, 1.1, and 2.0, among a sample of 27
long-period AGB variables. The fraction of stars with detectable Li line
(18.8\%) on the upper RGB ($J_0<12\fm0$) is similar to that found among AGB
stars ($4/27=14.8\%$). It is possible that the Li-rich AGB stars inherited their
Li from the preceding RGB phase. Indeed, the three Li-rich stars identified in
\citet {Leb12} could be early AGB stars instead of RGB stars because they are
brighter than the RC. Because it is impossible to separate RGB and early AGB
stars by means of our photometric and spectroscopic data alone,
asteroseismological methods would be needed to define their precise evolutionary
state.

\begin{figure}[t!]
\resizebox{\hsize}{!}{\includegraphics[clip=true]{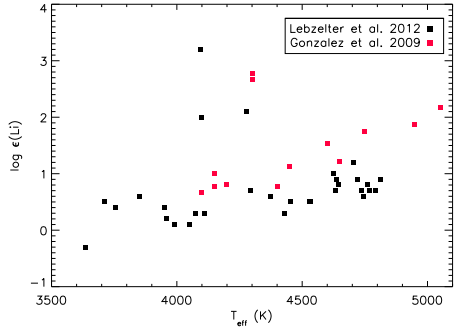}}
\caption{\footnotesize
Li abundance ($\log\epsilon({\rm Li})$) versus effective temperature of the
sample stars. Black symbols are stars from the survey of \citet{Leb12}, red
symbols are stars from the study of \citet{Gon09}. The Li-rich stars deviate
from the trend of decreasing Li abundance with decreasing temperature defined
by the Li-{\em detected} stars.}
\label{ALi_Teff}
\end{figure}

Another interesting result of our observations is that there is a trend for the
Li-{\em detected} stars of decreasing Li abundance with decreasing temperature,
from which the Li-{\em rich} stars deviate (Fig.~\ref{ALi_Teff}). This is in
qualitative agreement with what \citet{Gon09} found in their sample, our sample
extends this trend to lower temperatures. At a given temperature, our
Li abundances are on average somewhat lower than those of the \citet{Gon09}
sample. As elaborated on by them, the trend is not an artefact due to a
correlation between temperature determination and abundance measurement.
Furthermore, it is not connected to the decrease of the Li detection threshold
with decreasing temperature, as most of the Li abundances are clearly above the
thresold. This trend of decreasing Li abundance may be understood in the contect
of the parametric mixing models presented in \citet[][see their Fig.~1]{Pal11}.

%+++++++++++++++++++++++++++++++++++++++++++++++++++++++++++++++++++++++++++++++
%+++++++++++++++++++++++++++++++++++++++++++++++++++++++++++++++++++++++++++++++

\subsection{Mass loss}\label{mass_loss}

There are claims in the literature that the Li-rich giant phenomenon could be
accompanied by and connected with an episode of enhanced mass loss
\citep{dlR96,dlR97}. We therefore searched for indicators of mass loss from our
Li-rich sample stars. A comparison of the profile of the H$\alpha$ line of
Li-rich and Li-poor stars showed that there are no asymmetries in the lines
of the Li-rich stars. This suggests that the {\em gas} mass-loss rate from these
stars is not enhanced, within the resolution and S/N limits of our spectra.

\begin{figure}[t!]
\resizebox{\hsize}{!}{\includegraphics[clip=true]{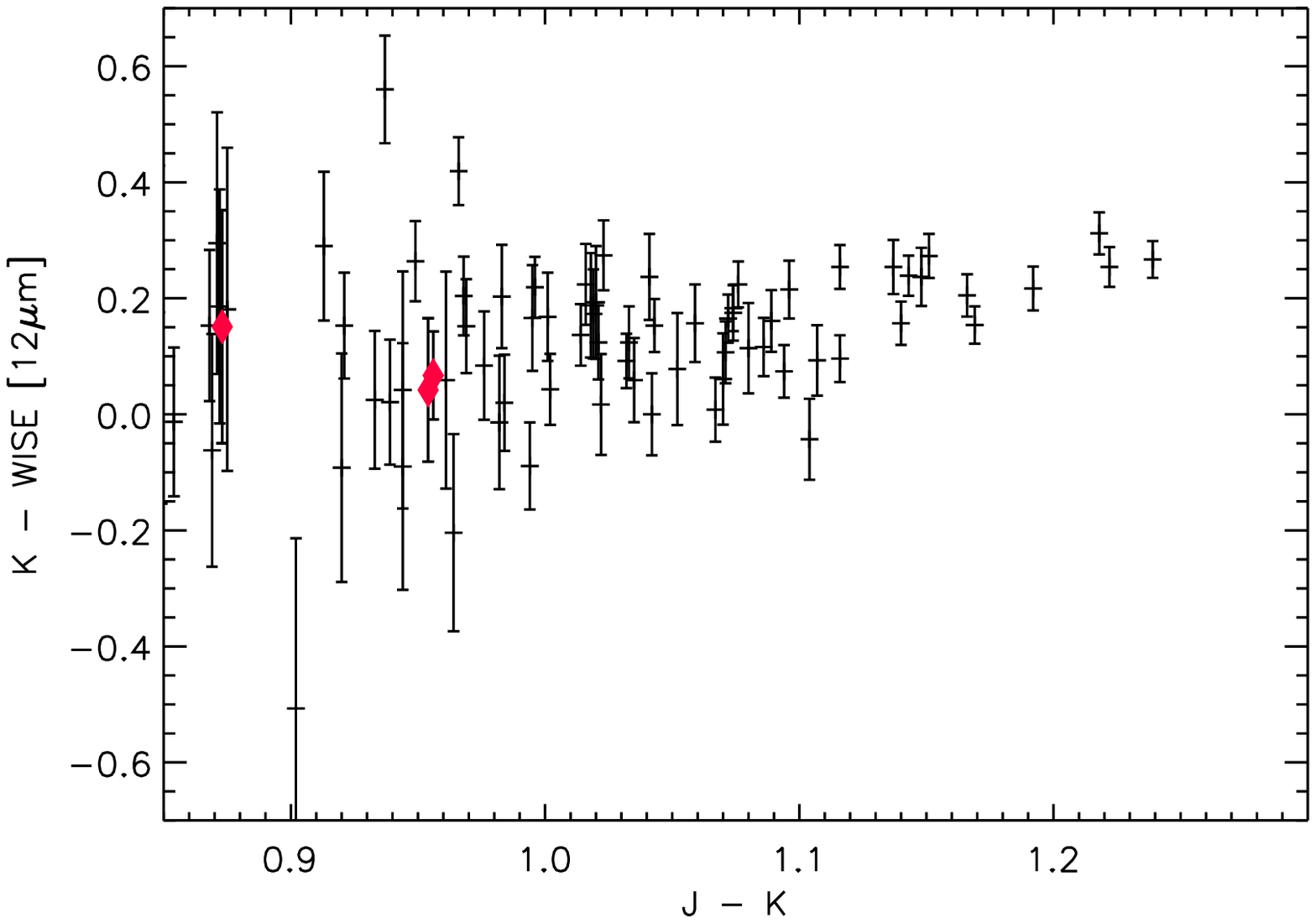}}
\resizebox{\hsize}{!}{\includegraphics[clip=true]{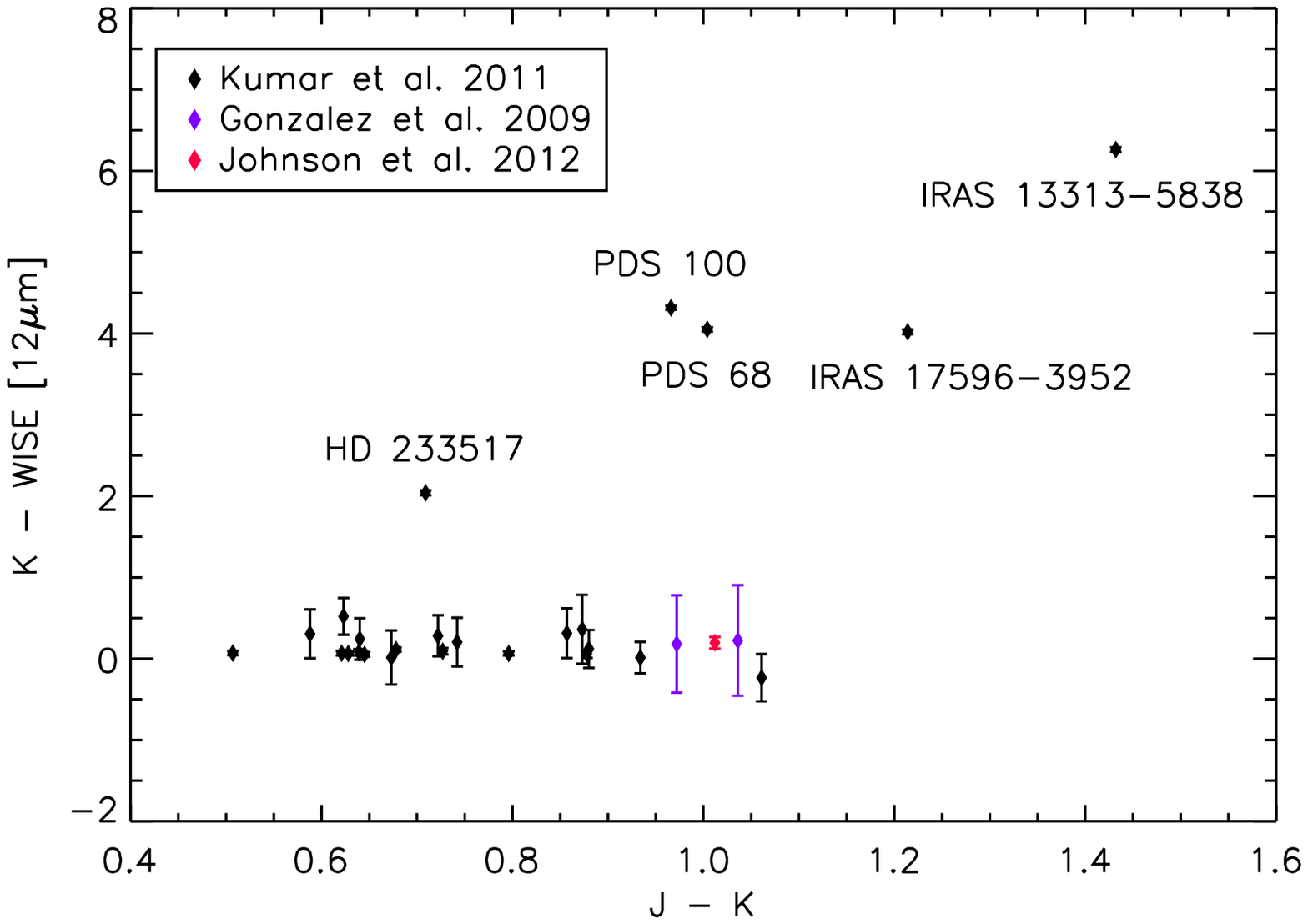}}
\caption{\footnotesize
{\em Upper panel:} $K - [12]$ vs.\ $J-K$ colour-colour diagram for the sample of
\citet{Leb12}. The Li-rich stars are represented by red diamond symbols.
{\em Lower panel:} The same diagram for Li-rich stars from the literature, as
indicated in the legend. The marked objects are discussed in the text
(Sect.~\ref{mass_loss}). Note the change in scales between the panels.}
\label{K12_lit}
\end{figure}

An excess of flux in the mid-IR is an indicator for {\em dusty} mass loss. We
therefore searched the Wide-field Infrared Survey Explorer (WISE) point source
catalogue\footnote{http://irsa.ipac.caltech.edu/} for counterparts of our
Li-rich stars. The $K-[12\mu{\rm m}]$ colour is particularly sensitive to warm
dust around stars. The result for our sample stars is shown in the upper panel
of Fig.~\ref{K12_lit}. The Li-rich stars, represented by red diamond
symbols, have a $K - [12]$ colour consistent with zero, i.e.\ no enhanced dust
mass-loss rate.

%+++++++++++++++++++++++++++++++++++++++++++++++++++++++++++++++++++++++++++++++
%+++++++++++++++++++++++++++++++++++++++++++++++++++++++++++++++++++++++++++++++

We also searched for the WISE 12\,$\mu$m fluxes of Li-rich stars reported in
the literature to compile a similar diagram for them. Most of these literature
data come from Tables~1 and 2 of \citet{Kumar11}, where Table~2 is itself a
collection of literature data (see their paper for references), but also the
Li-rich stars reported by \citet{Gon09} and \citet{John12} are included. The
release of the WISE all-sky catalogue has to be awaited to also include the
Li-rich stars from other studies \citep[e.g.\ ][]{Mon11,Ruch11}. The result of
this exercise is displayed in the lower panel of Fig.~\ref{K12_lit}. It is clear
that most stars form a sequence of photospheric colours ($K-[12]\approx0$), only
few stars deviate from this sequence, most of which are not normal K giants. For
instance, \object{IRAS 13313-5838} is a well-known post-AGB star. It is possible
that its strong 671\,nm line is actually not caused by Li, but rather by Ce
\citep{Ren02}. Also \object{IRAS 17596-3952} is a post-AGB candidate. Both
\object{PDS 68} and \object{PDS 100} (aka \object{V859 Aql}) are probably
T~Tauri stars. In that case they would still carry their primordial Li
abundance in the photosphere, which naturally is higher than the Li abundance
expected for K giants. Only \object{HD 233517} is a K-type giant with a dust
disk of unknown origin around it.

Thus, we think that warm dust ($\sim300$\,K) is not generally present around
Li-rich stars. However, Kumar et al.\ (this issue) find that the mass-loss rate,
as infered from IRAS photometry, correlates with Li abundance, albeit with a
large scatter. We cannot exclude the presence of cool dust by investigating the
$K-[12]$ colour, however the fraction of Li-rich stars with a considerable
amount of warm dust around them must be low.

\section{Correlation between lithium and technetium in low-mass AGB stars}

Technetium (Tc) is an element that has only radioactively unstable isotopes, its
logest-lived isotope produced by the s-process in low-mass AGB stars is
$^{99}$Tc with $\tau_{1/2}\approx2.1\times10^5$\,yrs. If absorption lines of this
element are identified, it is a clear sign of recent or ongoing s-process close
to the core of the star, and of a deep-mixing phenomenon called the third
dredge-up (3DUP). We refer to \citet{Utt07a} and references therein for details
on the s-process and 3DUP, as well as on previous observational results. Here
we concentrate on the correlation between the presence of Li and Tc in the
atmospheres of low-mass AGB stars.

Two of the Li-detected bulge AGB stars of \citet{Utt07} also show Tc in their
spectra (3DUP occurring), while two other stars showing Li are Tc-poor (no
3DUP). Furthermore, there are two stars in this sample that do have Tc, but no
Li. In this bulge sample, Tc is found only in stars with long pulsation periods
($P\gtrsim300$\,d) and high luminosities. Also Li is confined to the more
luminous and long-period objects ($P\gtrsim280$\,d). This already shows that
there must be some connection between the presence of Li and Tc in low-mass AGB
stars, although there is no 1:1 correlation. The high abundance of Li in one of
the Tc-poor bulge AGB stars was explained in \citet{Utt07} by a fast mixing
process below the convective envelope (``cool bottom processing'') that drives
the Cameron-Fowler mechanism.

A sample of 26 oxygen-rich (C/O$<$1), putative low-mass AGB stars in the
Galactic disc is available from \citet{UL10} and \citet{Utt11}. Also in this
combined sample, the fraction of Li-detected stars is higher among the stars
with Tc than among the stars without Tc (80.0\% compared with 43.8\%).
Unfortunately, the luminosities of Galactic disc stars are not well constrained,
but also in this sample the tendency for a star to have Li is higher at longer
pulsation period.

However, recent standard models of AGB evolution \citep{Kar10} suggest that the
observational picture could also be explained without invoking cool bottom
processes: $^7$Be-rich matter would be mixed to the surface naturally by 3DUP
events, where it decays to $^7$Li; the highest Li abundance would be found
already {\em before} efficient 3DUP of C and s-elements takes place
($\log\epsilon({\rm Li})=1.8$ in a $1.8M_{\sun}$ model). The Li- and Tc-poor
stars in the disc sample (9 out of 26) can then be interpreted as stars in which
the Li abundance was diminished in the previous stellar evolution to below the
detection threshold, and which did not (yet) undergo any ``Be dredge-up''. In
the 7 stars with Li but without Tc, Be-rich matter was already dredged-up from
regions close to the H-burning shell, but the convective envelope did not yet
reach the regions where the s-process takes place. Only stars in the more
advanced stages, which show both Li and Tc (8 out of 26 in this disc sample),
dredge up also s-process enriched matter besides $^7$Be. This evolution is
supported by the fact that most stars seem to follow the evolution from
(Li-poor,Tc-poor) $\rightarrow$ (Li-rich,Tc-poor) $\rightarrow$
(Li-rich,Tc-rich). Only two stars in this sample do have Tc but no Li, which
poses a small problem to this theoretical framework. Either these stars have
already lost all their $^3$He, or the Li is destroyed by slow mixing in the
inter-pulse phase between the 3DUP events.

The correlation between Li and Tc was also studied among S-stars (C/O$\sim$1) by
\citet{Van07}. However, the situation is more complex there because of extrinsic
S-stars (whose s-elements come from binary mass transfer), and because some
stars might produce their large amount of Li by hot bottom burning.

\section{Lithium production by hot bottom burning in long-period Miras}

Bright red giant stars with very strong Li lines have been detected in the
Magellanic clouds \citep{Smith95}, but also in the Milky Way galaxy
\citep{GH07}. These are interpreted as being intermediate-mass stars
($M\gtrsim4M_{\sun}$) in which the H-burning takes place under convective
conditions, called ``hot bottom burning'' (HBB). This allows for the
Cameron-Fowler mechanism to produce large amounts of Li.

In our study of the evolutionary state of Miras with changing pulsation periods
\citep{Utt11}, we also identified stars that have a very strong Li 671\,nm line
and are thus candidates for intermediate-mass AGB stars undergoing HBB. The best
example for this is the long-period Mira \object{R Nor} ($P\sim500$\,d). Besides
the high Li abundance \citep[$\log\epsilon({\rm Li})=4.6$;][]{Utt11}, indicators
of a high mass are the secondary maximum in its light curve, the high
luminosity, and the relatively small distance to the Galactic plane.
Interestingly, R~Nor was found to be Tc-poor. This could be a sign of either low
dredge-up efficiency in intermediate-mass AGB stars, or the operation of the
$^{22}$Ne neutron source, which would hardly produce any Tc.

\begin{figure}[t!]
\resizebox{\hsize}{!}{\includegraphics[clip=true]{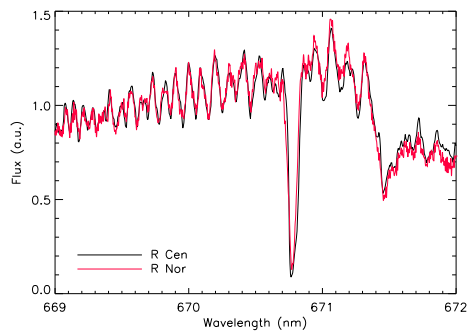}}
\caption{\footnotesize
The spectra of \object{R Cen} and \object{R Nor} around the Li 671\,nm line.}
\label{RCenRNor}
\end{figure}

In our search for more Miras that could be intermediate-mass AGB stars, we found
a UVES spectrum of \object{R Cen}, a long-period Mira that is also known to
change its pulsation period \citep{Haw01}. Figure~\ref{RCenRNor} shows a
comparison between the spectrum of R~Nor and R~Cen around the Li 671\,nm line.
The spectra are remarkably similar, even though the stars were observed with
two different instruments, approximately ten years apart. Thus, we may assume
that the stars have very similar Li abundance. Just as R~Nor, R~Cen has a
period of $P\sim500$\,d (although the period may have halved in the recent years
because the primary minima became much shallower) and is at a very small
distance to the Galactic plane (only 8\,pc), all of which are indicators of a
relatively high mass of R~Cen ($M\sim4M_{\sun}$). Also R~Cen has no Tc.

%+++++++++++++++++++++++++++++++++++++++++++++++++++++++++++++++++++++++++++++++
%+++++++++++++++++++++++++++++++++++++++++++++++++++++++++++++++++++++++++++++++

A number of atomic resonance lines in both R~Nor and R~Cen (Na\,D doublet, Li
6708\,\AA, K 7700\,\AA) show blue-shifted absorption components from
circumstellar material, indicating that these stars are losing mass. Thus, they
might be in a phase where they pollute the interstellar matter with Li-rich
material.

\section{Conclusions and outlook}

The phenomenon of the Li-rich K giants is still mysterious. Despite more and
more such stars are being found in diverse stellar systems, no satisfactory
explanation can be given for them. A step forward would be to determine precise
luminosities, for which precise distances are required. Parallax measurements of
many such stars will become available in the Gaia era. Furthermore, it would be
of great help to know whether these stars are preferentially H-shell burning RGB
or He-burning early AGB stars. The tools of asteroseismology can distinguish
between these evolutionary stages
\citep{Bed11}.

The Li-rich long-period Miras should be studied in more detail in the future.
Only few observational constraints of the evolution of intermediate-mass stars
on the AGB are available at the moment. However, they potentially play an
important role in the evolution of Li in the Galaxy and in star clusters
because of their very high Li abundance and high mass-loss rate. They might be
responsible for polluting the intra-cluster gas of globular clusters
\citep{Lind09}, thus they are of importance on a larger scale.

\begin{acknowledgements}
We thank the organisers for giving us the opportunity to present our work at
this workshop. SU acknowledges support from the Austrian Science Fund (FWF)
under project P~22911-N16, and TL under projects P~21988-N16 and P~23737-N16.
%BA thanks for the support from contract ASI-INAF I/009/10/0.
\end{acknowledgements}

\bibliographystyle{aa}

\end{document}